\documentclass[prl, twocolumn]{revtex4}

\usepackage{amsmath,amssymb,amscd}  
\usepackage{psfrag}  
\usepackage[dvips]{epsfig}  
\usepackage{epsfig}

\usepackage{enumerate}

\usepackage{theorem}
\theorembodyfont{\rmfamily}

\theoremstyle{break}

\def\QED{~\rule[-1pt]{5pt}{5pt}\par\medskip}

\def\k{\mathfrak{k}}

\def\p{\mathfrak{p}}
\def\k{\mathfrak{k}}

\def\su{\mathfrak{su}}

\def\Eq{Eq.~\eqref}
\def\Eqs{Eqs.~\eqref}

\DeclareMathOperator{\diag}{diag}

\DeclareMathOperator{\Ent}{Ent}

\newcommand{\ma}[1]{\left[\begin{matrix} #1 \end{matrix}\right]}

\def\ie{{\it i.e.}}

\begin{document}   

\title{Time evolution of Two-qubit Entanglement}

\author{Jun Zhang} \affiliation{Department of Automation, Joint
  Institute of UMich-SJTU, Shanghai Jiao Tong University;\\
  Key Laboratory of System Control and Information Processing,
  Ministry of Education, Shanghai, China} 

\date{May 26, 2011}
  
\begin{abstract} 
  We show that the entanglement dynamics for a closed two-qubit system
  is part of a 10-dimensional complex linear differential equation
  defined on a supersphere, and the coefficients therein are
  completely determined by the Hamiltonian. We apply the result to
  investigate two physical examples of Josephson junction qubits and
  exchange Hamiltonians, deriving analytic solutions for the time
  evolution of entanglement. The Hamiltonian coefficients determine
  whether the entanglement is periodic. These results allow of
  investigating how to generate and manipulate entanglements
  efficiently, which are required by both quantum computation and
  quantum communication.
\end{abstract}   
  
\maketitle 

Entanglement is one of the most striking quantum mechanical properties
that plays a central role in quantum computation and quantum
communication~\cite{Nielsen:00}. It is a major resource used in many
applications such as quantum algorithm, teleportation, and quantum
cryptography. In recent years, there are several problems receiving
considerable research efforts, including dynamical evolution of
entanglement where the objective is to find out how the entanglement
of a quantum system evolves as time elapses~\cite{Mintert:05,
  Horodecki:09}. This is especially important to quantum information
processing that relies primarily upon the generation, manipulation,
and detection of quantum entanglement.

Many of the current researches on this topic focus on the entanglement
decay or production for an open quantum system interacting with the
surrounding environment. For example, Konrad {\it et
  al.}~\cite{Konrad:08} proved a factorization law for bipartite
system that describes entanglement evolutions with a noisy channel.
Yu and Eberly~\cite{Yu:09} revealed that quantum entanglement
influenced by environmental noise can undergo a sudden death. See
Refs.~\cite{Zycz:01, Dodd:04, Dur:04, Santos:06, Carvalho:07, Roos:08,
  Konrad:08, Yu:09} and the references therein for full details.

Here we consider a different problem: for a given closed two-qubit
quantum system with Hamiltonian $H$, what are the dynamics and the
resulting time evolution of its entanglement? To our knowledge, this
is a basic yet to be answered question. In many applications of
quantum computation and quantum communication, often desired is to
efficiently generate entanglement from some initial state.  Therefore,
it is of particular value to investigate how the entanglement of a
qubit system evolves as a function of time so as to analyze the
capability of a quantum system to produce and further manipulate
quantum entanglement. Moreover, it will also help us to have a deeper
understanding of the fundamentals of quantum physics.

We will show that for a closed two-qubit system, its entanglement
dynamics can be described by part of a 10-dimensional complex
differential equation. All the coefficients in this equation are
determined by the Hamiltonian. For specific two-qubit systems of
Josephson junction and exchange Hamiltonians, we will derive
closed-form solutions for the time evolution of entanglement.

First we briefly introduce some backgrounds (See Ref.~\cite{Zhang:02}
for details). The quantum operations for a two-qubit system are
defined on the special unitary Lie group $SU(4)$.  The associated Lie
algebra is denoted as $\su(4)$ and has a direct sum decomposition
$\su(4)=\p\oplus\k$, where
\begin{equation}
  \label{eq:29}
  \begin{split}
    \k &= \text{span } \frac{i}2\{\sigma_x^1,\ \sigma_y^1,\  \sigma_z^1,\
  \sigma_x^2,\  \sigma_y^2,\  \sigma_z^2\},\\
\p &=\text{span } \frac{i}2\{\sigma_x^1\sigma_x^2,\  \sigma_x^1\sigma_y^2,\
\sigma_x^1\sigma_z^2,\  \sigma_y^1\sigma_x^2,\  \sigma_y^1\sigma_y^2,\\
&\quad\quad \sigma_y^1\sigma_z^2,\  \sigma_z^1\sigma_x^2,\ \sigma_z^1\sigma_y^2,\ 
\sigma_z^1\sigma_z^2 \}.
  \end{split}
\end{equation}
Here $\sigma_x$, $\sigma_y$, and $\sigma_z$ are the Pauli matrices,
and $\sigma_{\alpha}^1\sigma_{\beta}^2 = \sigma_{\alpha} \otimes
\sigma_{\beta}$. The set $\k$ contains all the local terms, whereas
$\p$ has the nonlocal or coupling terms. An arbitrary Hamiltonian for
two-qubit system can be represented by a linear combination of the
basis matrices in \Eq{eq:29} as
\begin{eqnarray}
  \label{eq:25}
& &H=\frac{a_1}2\sigma_x^1 +\frac{a_2}2\sigma_y^1 +\frac{a_3}2\sigma_z^1
 +\frac{a_4}2\sigma_x^2+\frac{a_5}2\sigma_y^2 +\frac{a_6}2\sigma_z^2\nonumber \\
&&\ +\frac{a_7}2\sigma_x^1\sigma_x^2+\frac{a_8}2\sigma_x^1\sigma_y^2+ 
 \frac{a_9}2\sigma_x^1\sigma_z^2+\frac{a_{10}}2\sigma_y^1\sigma_x^2+
 \frac{a_{11}}2\sigma_y^1\sigma_y^2\nonumber \\
&&\ +\frac{a_{12}}2\sigma_y^1\sigma_z^2
 +\frac{a_{13}}2\sigma_z^1\sigma_x^2+\frac{a_{14}}2\sigma_z^1\sigma_y^2+  
 \frac{a_{15}}2\sigma_z^1\sigma_z^2,
\end{eqnarray}
where $a_k$'s are real numbers.  Denote the state of the quantum
system as $\psi$.  The dynamics of $\psi$ is determined by
Schr\"odinger's equation:
\begin{equation}
  \label{eq:1}
  \dot\psi  = iH\psi
\end{equation}
with initial state $\psi(t_0)$, where $t_0$ is the initial time.

We use the concurrence of $\psi$ as an entanglement measure, which is
defined in Refs.~\cite{Wootters:98, Makhlin:00s} as
\begin{equation*}
  C(\psi)=| \Ent \psi |,
\end{equation*}
where $\Ent{\psi}=\psi^T \sigma_y^1\sigma_y^2\,\psi$. It can be shown
that the concurrence $C(\psi)$ is invariant under the local operations
and it ranges from $0$ to $1$.  The condition $C(\psi)=0$ holds true
if and only if ${\psi}$ is an unentangled state. In the case when
$C(\psi)$ achieves maximal value $1$, such ${\psi}$ is called a
maximally entangled state. We have previously given the conditions
for $\psi$ to be maximally entangled~\cite{Zhang:02}.

The concurrence $C(\psi)$ defines a measure of entanglement for the
two-qubit pure state ${\psi}$.  In what follows, we will derive the
dynamics of $\Ent\psi$, that is, the differential equation that
governs its time evolution. To this end, take derivative of $\Ent
\psi$:
\begin{eqnarray}
  \label{eq:30}
&&\frac{d}{dt} \Ent \psi =\dot \psi^T \sigma_y^1\sigma_y^2\psi
+\psi^T \sigma_y^1\sigma_y^2 \dot\psi\nonumber\\
&=&i\psi^T (H^T\sigma_y^1\sigma_y^2
+ \sigma_y^1\sigma_y^2 H)\psi\nonumber\\
&=& i\psi^T \big(a_{11}I- a_{15}\sigma_x^1\sigma_x^2
- a_{7}\sigma_z^1\sigma_z^2
- a_{8}i\sigma_z^1   
+ a_{12} i \sigma_x^2  \nonumber\\
&&+ a_{13}  \sigma_x^1\sigma_z^2 
+ a_{14}   i\sigma_x^1 
+ a_{9}  \sigma_z^1\sigma_x^2
- a_{10}i \sigma_z^2\big)\psi.
\end{eqnarray}
The derivative of $\Ent\psi$ depends on the terms such as $\psi^T\psi$
and $\psi^T\sigma_x^1\sigma_x^2\psi$ in \Eq{eq:30}. To get the
complete dynamics of $\Ent\psi$, we need to find the derivatives of
all these terms. For the ease of notation, define
\begin{equation}
  \label{eq:14}
x_k=\psi^TP_k\psi, \quad k=1, \dots, 10,
\end{equation}
where
\begin{equation}
  \label{eq:24}
  \begin{aligned}
 P_1&=\sigma_y^1\sigma_y^2, &P_2&=I,&
 P_3&=\sigma_x^1\sigma_x^2,\\
 P_{4}&=\sigma_z^1\sigma_z^2,&
 P_{5}&=i\sigma_z^1, &P_{6}&=i \sigma_x^2,\\
 P_{7}&=\sigma_x^1\sigma_z^2,&
 P_{8}&=i\sigma_x^1,&
 P_{9}&=\sigma_z^1\sigma_x^2,\quad
 P_{10}=i \sigma_z^2.
  \end{aligned}
\end{equation}
It is clear that $x_1=\Ent\psi$, and \Eq{eq:30} can be rewritten as
\begin{eqnarray*}
  \label{eq:32}
\dot x_1&=& i (a_{11}x_2 - a_{15}x_3- a_{7}x_4
- a_{8}x_5+ a_{12}x_6\nonumber\\
&&+ a_{13}x_7+a_{14}x_8+a_{9}x_9-a_{10}x_{10}).
\end{eqnarray*}
For the other $x_k$'s, we similarly obtain
\begin{equation}
  \label{eq:33}
 \dot x_k=\dot\psi^TP_k\psi +\psi^TP_k\dot\psi
=i\psi^T(H^TP_k+P_kH)\psi.
\end{equation}
One salient feature of the matrices $\{P_j\}_{j=1}^{10}$ defined in
\Eq{eq:24} is that any matrix in the form of $H^TP_k+P_kH$ can be
represented by a linear combination of all these matrices. Therefore,
$\dot x_k$ in \Eq{eq:33} can be written as a linear combination of all
the $x_j$'s.

Let $x=[x_1, \dots, x_{10}]^T$. After some derivations, we obtain that
$x$ satisfies the following linear differential equation
\begin{equation}
  \label{eq:12}
  \dot x=iAx,
\end{equation}
where 
\begin{widetext}
\begin{equation}
  \label{eq:15}
A=\left[\begin{array}{c|c|c}
A_{11}&A_{12}&A_{13}\\ \hline A_{12}^\dag&A_{22}&A_{23}\\ \hline
 A_{13}^\dag&A_{23}^\dag&A_{33}  \end{array}\right]=
\left[
  \begin{array}{cccc|ccc|ccc}
0&a_{11}&-a_{15}&-a_{7}&-a_{8}&a_{12}&a_{13}&a_{14}&a_{9}&-a_{10}\\
a_{11}&0&a_{7}&a_{15}&-ia_{3}&-ia_{4}&a_{9}&-ia_{1}&a_{13}&-ia_{6}\\
-a_{15}&a_{7}&0&-a_{11}&a_{10}&-ia_{1}&ia_{5}&-ia_{4}&ia_{2}&a_{8}\\
-a_{7}&a_{15}&-a_{11}&0&-ia_{6}&-a_{14}&-ia_{2}&-a_{12}&-ia_{5}&-ia_{3}\\ \hline
-a_{8}&ia_{3}&a_{10}&ia_{6}&0&a_{13}&a_{12}&-ia_{2}&ia_{4}&a_{15}\\
a_{12}&ia_{4}&ia_{1}&-a_{14}&a_{13}&0&-a_{8}&a_{7}&ia_{3}&ia_{5}\\
a_{13}&a_{9}&-ia_{5}&ia_{2}&a_{12}&-a_{8}&0&-ia_{6}&a_{11}&-ia_{1}\\ \hline
a_{14}&ia_{1}&ia_{4}&-a_{12}&ia_{2}&a_{7}&ia_{6}&0&-a_{10}&a_{9}\\
a_{9}&a_{13}&-ia_{2}&ia_{5}&-ia_{4}&-ia_{3}&a_{11}&-a_{10}&0&a_{14}\\
-a_{10}&ia_{6}&a_{8}&ia_{3}&a_{15}&-ia_{5}&ia_{1}&a_{9}&a_{14}&0
  \end{array}\right].
\end{equation}
\end{widetext}
The block matrices $A_{ij}$'s in the first equality have a conforming
partition as those in the second equality. The elements in the $k$-th
row of $A$ are exactly those coefficients in the representation of
$H^TP_k+P_kH$ as a linear combination of $\{P_j\}_{j=1}^{10}$.
Eqs.~\eqref{eq:12} and~\eqref{eq:15} reveal that the entanglement
dynamics for a closed two-qubit system is part of a 10-dimensional
complex linear differential equations. It remains unknown to us why
the dimension is $10$ though. We can also split the real and imaginary
parts of $x$ to get a real differential equation with dimension $20$.
This is the key result of this paper.

The matrix $A$ is Hermitian, \ie, $A=A^\dag$. Then,
\begin{equation*}
  \label{eq:10}
\frac{d}{dt}x^\dag x
=\dot x^\dag x+x^\dag\dot x
=-i x^\dag A^\dag x+x^\dag iA x=0,
\end{equation*}
which yields that
\begin{equation*}
  \label{eq:35}
  \|x(t)\|^2=x^\dag(t) x(t)=x^\dag(t_0) x(t_0),
\end{equation*}
for all $t\ge t_0$. This avers that the norm of $x(t)$ is conserved
along the trajectory of Schr\"odinger's equation~\eqref{eq:1}.  We now
show that this conserved quantity is $2$. Let the initial state of
$\psi$ be
\begin{equation}
  \label{eq:6}
 \psi(t_0)=[\psi_1+i\psi_2, \psi_3+i\psi_4, \psi_5+i\psi_6, \psi_7+i\psi_8]^T,
\end{equation}
where $\psi_l$'s are real numbers and satisfy $\sum_{l=1}^8 \psi_l^2=1$.
Let the initial condition of $x_k$ be 
\begin{equation}
  \label{eq:16}
  x_k(t_0)=p_k+i q_k,
\end{equation}
\ie, $p_k$ and $q_k$ are the real and imaginary parts of $x_k(t_0)$,
respectively. Substituting \Eq{eq:6} into \Eq{eq:14}, we can represent
$p_k$ and $q_k$ in terms of $\psi_l$'s as
\begin{equation}
  \label{eq:19}
  \begin{split}
p_1&=2(\psi_3 \psi_5-\psi_4 \psi_6-\psi_1 \psi_7+\psi_2 \psi_8),\\
p_2&=\psi_1^2-\psi_2^2+\psi_3^2-\psi_4^2+\psi_5^2-\psi_6^2+\psi_7^2-\psi_8^2,\\
p_3&=2( \psi_3\psi_5-\psi_4 \psi_6+\psi_1 \psi_7-\psi_2 \psi_8),\\
p_4&=\psi_1^2-\psi_2^2-\psi_3^2+\psi_4^2-\psi_5^2+\psi_6^2+\psi_7^2-\psi_8^2,\\
p_5&=2(-\psi_1\psi_2-\psi_3 \psi_4+\psi_5 \psi_6+\psi_7 \psi_8),\\
p_6&=2(-\psi_2 \psi_3-\psi_1 \psi_4-\psi_6 \psi_7-\psi_5 \psi_8),\\
p_7&=2(\psi_1\psi_5-\psi_2 \psi_6-\psi_3 \psi_7+\psi_4 \psi_8),\\
p_8&=2(-\psi_2 \psi_5-\psi_1 \psi_6-\psi_4 \psi_7-\psi_3 \psi_8),\\
p_9&=2(\psi_1\psi_3-\psi_2 \psi_4-\psi_5 \psi_7+\psi_6 \psi_8),\\
p_{10}&=2(-\psi_1 \psi_2+\psi_3 \psi_4-\psi_5 \psi_6+\psi_7\psi_8),
  \end{split}
\end{equation}
and
\begin{equation}
  \label{eq:34}
  \begin{split}
q_1&=2(\psi_4 \psi_5+\psi_3 \psi_6-\psi_2 \psi_7-\psi_1 \psi_8),\\
q_2&=2(\psi_1 \psi_2+\psi_3 \psi_4+\psi_5 \psi_6+\psi_7 \psi_8),\\
q_3&=2(\psi_4 \psi_5+\psi_3 \psi_6+\psi_2 \psi_7+\psi_1 \psi_8),\\
q_4&=2(\psi_1 \psi_2-\psi_3 \psi_4-\psi_5 \psi_6+\psi_7 \psi_8),\\
q_5&=\psi_1^2-\psi_2^2+\psi_3^2-\psi_4^2-\psi_5^2+\psi_6^2-\psi_7^2+\psi_8^2,\\
q_6&=2(\psi_1 \psi_3-\psi_2 \psi_4+\psi_5 \psi_7-\psi_6 \psi_8),\\
q_7&=2(\psi_2 \psi_5+\psi_1 \psi_6-\psi_4 \psi_7-\psi_3 \psi_8),\\
q_8&=2(\psi_1 \psi_5-\psi_2 \psi_6+\psi_3 \psi_7-\psi_4 \psi_8),\\
q_9&=2(\psi_2 \psi_3+\psi_1 \psi_4-\psi_6 \psi_7-\psi_5 \psi_8),\\
q_{10}&=\psi_1^2-\psi_2^2-\psi_3^2+\psi_4^2+\psi_5^2-\psi_6^2-\psi_7^2+\psi_8^2.
  \end{split}
\end{equation}
Proceeding further, we have
\begin{eqnarray*}
  \label{eq:18}
\|x(t)\|=\sqrt{ x^\dag(t_0) x(t_0)}
=\sqrt{ \sum_{k=1}^{10} \left(p_k^2+q_k^2\right)}
=2\sum_{l=1}^8 \psi_l^2=2.
\end{eqnarray*}
Therefore, the dynamics of \Eq{eq:12} is defined on a supersphere with
radius 2.

For a general Hamiltonian in \Eq{eq:25}, there exists a local
operation $k\in SU(2)\otimes SU(2)$ such that all the coupling terms
in $kHk^\dag$ vanish except $\sigma_x^1\sigma_x^2$, $\sigma_y^1
\sigma_y^2$, and $\sigma_z^1 \sigma_z^2$~\cite{Zhang:02}:
 \begin{eqnarray}
 \label{eq:26}
kHk^\dag&=&\frac{a_1}2\sigma_x^1 +\frac{a_2}2\sigma_y^1 
+\frac{a_3}2\sigma_z^1+\frac{a_4}2\sigma_x^2 
+\frac{a_5}2\sigma_y^2+\frac{a_6}2\sigma_z^2\nonumber\\
&&+\frac{a_7}2\sigma_x^1\sigma_x^2+\frac{a_{11}}2\sigma_y^1\sigma_y^2 
+\frac{a_{15}}2\sigma_z^1\sigma_z^2.
 \end{eqnarray}
 By an abuse of notation, we again use $a_k$ to denote the
 coefficients in the right hand side of \Eq{eq:26}. Because
 \begin{equation*}
   \label{eq:37}
   e^{iHt}\psi(t_0) =k^\dag (e^{ikHk^\dag}) k\psi(t_0)
 \end{equation*}
 and the local operation $k^\dag$ does not change the entanglement of
 quantum state, the function $\Ent \psi$ generated by $H$ with
 initial state $\psi(t_0)$ is the same as that by $kHk^\dag$ with
 $k\psi(t_0)$.  We then only need to study a simplified differential
 equation, where all the entries in $A$ corresponding to
 cross-coupling terms are 0. In particular, two diagonal blocks
 $A_{22}$ and $A_{33}$ in \Eq{eq:15} both become zero matrices.
 
 To illustrate the idea, we now study the time evolution of
 entanglement for two physical examples, namely, charge-coupled
 Josephson junction and exchange Hamiltonians. We focus on the
 Hamiltonians rather than physical implementation details, because the
 Hamiltonian completely determines the entanglement dynamics.
 
 First consider a charge-coupled Josephson junction qubit system
 discussed in Ref.~\cite{Makhlin:99}. The Hamiltonian is given by $
 H_1=-( E_J/2)(\sigma_x^1+\sigma_x^2)+ ({E_J^2}/{E_L})
 \sigma_y^1\sigma_y^2$,
 which contains both local and nonlocal terms. Then $a_1=a_4=-E_J$
 and $a_{11}=2E_J^2/E_L$. Setting all the other $a_k$'s to 0 in
 \Eq{eq:15}, we obtain a reduced order differential equation:
\begin{equation*}
  \label{eq:38}
\left[\begin{matrix}
 \dot x_1\\\dot x_2 \\ \dot x_3 \\ \dot x_4 \\ \dot x_6 \\ \dot x_8   
  \end{matrix}\right]= i
\left[\begin{array}{ccc|ccc}
    0 & a_{11} & 0 & 0 &0 &0\\
  a_{11} & 0 & 0 & 0 &-ia_{1} &-ia_1\\
 0&0&0&- a_{11} & -ia_1 & -ia_1\\ \hline
 0&0&-a_{11}&0&0 &0\\ 
0& ia_{1} & ia_{1} & 0 &0 &0\\
0& ia_{1} & ia_{1} & 0 &0 &0
  \end{array}\right]
\left[\begin{matrix}
  x_1\\ x_2 \\ x_3 \\ x_4 \\ x_6 \\ x_8   
  \end{matrix}\right].
\end{equation*}
Let $\alpha=E_J/E_L$. Then $a_{11}=-2\alpha a_1$. Solving the
differential equation above yields
\begin{equation*}
  \label{eq:44}
  \Ent \psi(t)=x_1(t)=r_1(t)+is_1(t),
\end{equation*}
where 
\begin{eqnarray}
  \label{eq:4}
&&r_1(t)\nonumber\\
&=&\frac{(p_1-p_4)+(q_6+q_8)\alpha} {2(1+\alpha^2)}
+\frac{(q_2+q_3)\alpha}{2\sqrt{1+\alpha^2}}
\sin  \sqrt{1+\alpha^2} E_J t\nonumber \\
&+&\frac{(p_1-p_4)\alpha - (q_6+q_8)}{2(1+\alpha^2)}
\alpha \cos  \sqrt{1+\alpha^2} E_Jt\nonumber \\
&+&\frac{(p_1+p_4)}{2}\cos  \alpha E_Jt
+\frac{(q_2-q_3)}{2} \sin \alpha E_J t,
\end{eqnarray}
and
\begin{eqnarray}
  \label{eq:5}
&&s_1(t)\nonumber\\
&=&\frac{(q_1-q_4)-(p_6+p_8)\alpha}{2(1+\alpha^2)}
-\frac{(p_2+p_3)\alpha}{2 \sqrt{1+\alpha^2} }
 \sin \sqrt{1+\alpha^2} E_J t\nonumber \\
&+&\frac{(q_1-q_4)\alpha+(p_6+p_8)}{2(1+\alpha^2)}\alpha
\cos \sqrt{1+\alpha^2} E_J t \nonumber \\
&+&\frac{(q_1+q_4) }{2} \cos \alpha E_J t
-\frac{(p_2-p_3)}{2}\sin \alpha E_J t.
\end{eqnarray}
Here $p_k$'s and $q_k$'s are defined in \Eqs{eq:19} and~\eqref{eq:34}.
Hence,
\begin{equation*}
  \label{eq:46}
C(\psi)={\sqrt{r_1^2(t)+s_1^2(t)}}.
\end{equation*}
The entanglement evolution has two frequency components
$\sqrt{1+\alpha^2} E_J$ and $\alpha E_J$. When the ratio between these
two values,$\sqrt{1+\alpha^2}/\alpha$, is a rational number,
entanglement is periodic; otherwise, it is aperiodic.

Next let us consider two-qubit exchange Hamiltonian $H_2 =
\frac12({a_7}\sigma_x^1\sigma_x^2+ a_{11} \sigma_y^1\sigma_y^2 +
a_{15}\sigma_z^1\sigma_z^2)$.
In this case, all the local terms vanish, \ie, $a_1=0$, \dots,
$a_6=0$. For the block matrices in \Eq{eq:15}, we have
$A_{12}=A_{13}=0$. Therefore, the dynamics of $x_u=[x_1, x_2, x_3,
x_4]^T$ is decoupled from that of $[x_5, \dots, x_{10}]^T$, which
leads to
\begin{equation}
  \label{eq:45}
  \dot x_u=i A_{11} x_u, 
\end{equation}
or more explicitly,
\begin{equation}
  \label{eq:3}
\ma{\dot x_1\\\dot x_2 \\ \dot x_3 \\ \dot x_4}
= i \ma{ 0 & a_{11} & -a_{15} & -a_7 \\
 a_{11} & 0 & a_7 & a_{15} \\
 -a_{15} & a_7 & 0 & -a_{11} \\
 -a_7 & a_{15} & -a_{11} & 0}
\ma{x_1\\ x_2 \\ x_3 \\ x_4}.
\end{equation}
Because $A_{11}$ is also a Hermitian (or symmetric indeed) matrix, the
norm of $x_u$ is also conserved. Let 
\begin{equation*}
  \label{eq:2}
T=\frac12\ma{-1&1&1&1\\1&-1&1&1\\1&1&-1&1\\1&1&1&-1}.
\end{equation*}
The matrix $A_{11}$ can be diagonalized as
\begin{eqnarray}
  \label{eq:28}
T^{-1}A_{11}T&=&\diag\{ a_7-a_{11}+a_{15}, \ -a_7-a_{11}-a_{15},\nonumber \\
&&-a_7+a_{11}+a_{15},\ a_7+a_{11}-a_{15} \}.
\end{eqnarray}
The entanglement evolution therefore has four frequency components as
given in \Eq{eq:28}. We can obtain the same diagonal matrix if
transforming the Hamiltonian $H_2$ into the Bell basis. We then have
\begin{equation*}
  \label{eq:11}
  \Ent \psi(t)=x_1(t)=\left(r^T(t) +is^T(t)\right) l,
\end{equation*}
where $l=[p_1,p_2,p_3,p_4,q_1,q_2,q_3,q_4]^T$, and
\begin{alignat*}{2}
r^T(t)=[
&\cos a_{7} t  \cos a_{11} t  \cos a_{15} t,&
&  \sin a_{7} t  \cos a_{11} t  \sin a_{15} t, \\
-&\sin a_{7} t  \sin a_{11} t  \cos a_{15} t,&
-&\cos a_{7} t  \sin a_{11} t  \sin a_{15} t,\\
&\sin a_{7} t  \sin a_{11} t  \sin a_{15} t,&
-&\cos a_{7} t  \sin a_{11} t  \cos a_{15} t,\\
&\cos a_{7} t  \cos a_{11} t  \sin a_{15} t,&
&\sin a_{7} t  \cos a_{11} t  \cos a_{15} t],
\end{alignat*}
\begin{alignat*}{2}
s^T(t)=[
&  \sin a_{7} t \sin a_{11} t  \sin a_{15} t ,& &
  \cos a_{7} t  \sin a_{11} t  \cos a_{15} t , \\
-&\cos a_{7} t  \cos a_{11} t  \sin a_{15} t , & 
-&\sin a_{7} t  \cos a_{11} t  \cos a_{15} t, \\
& \cos a_{7} t  \cos a_{11} t  \cos a_{15} t, & &
  \sin a_{7}t   \cos a_{11} t  \sin a_{15} t, \\
-&\sin a_{7} t  \sin a_{11} t  \cos a_{15} t, & 
-&\cos a_{7} t  \sin a_{11} t  \sin a_{15} t].   
\end{alignat*}
Therefore,
\begin{equation*}
  \label{eq:27}
C(\psi)=\sqrt{l^T\left(r(t)r^T(t)+s(t)s^T(t)\right)l}.
\end{equation*}
Examining the frequency components in \Eq{eq:28}, we know that if the
pairwise ratios between $a_7$, $a_{11}$, and $a_{15}$ are all rational
numbers, the entanglement measure is a periodic function.

For specific exchange Hamiltonian, we can further simplify the
entanglement measure. For instance, for two-dimensional XY exchange
Hamiltonian $H_3=\sigma_x^1\sigma_x^2 + \sigma_y^1\sigma_y^2$, we have
\begin{eqnarray*}
  \label{eq:13}
&C^2(\psi)
=\dfrac{1}{4} \left( (p_2-p_4)\sin 2t+ (q_1+q_3)\cos
  2t+q_1-q_3\right)^2 \nonumber\\
&+\dfrac14\left((q_2-q_4)\sin 2t-(p_1+p_3)\cos 2t-p_1+p_3\right)^2.
\end{eqnarray*}
For Ising Hamiltonian $H_4=\sigma_x^1\sigma_x^2$, we have
\begin{equation*}
  \begin{split}
C(\psi)=\sqrt{\left(p_4\sin t-q_1 \cos t\right)^2
+\left(q_4 \sin t+p_1\cos t\right)^2}. 
  \end{split}
\end{equation*}
It is clear that in these two cases, the concurrence measure $C(\psi)$
are both periodic functions.

In summary, we have derived the dynamical equation that governs the
time evolution of entanglement for closed two-qubit systems. This
turns out to be a 10-dimensional differential equation defined on a
supersphere.  We applied the result to investigate two physical
applications, namely, Josephson junction and exchange Hamiltonians.
For both cases, we derived analytic solutions for the concurrence
measure of entanglement. The coefficients in the Hamiltonian
completely determine whether or not the entanglement is a periodic
function. We expect to extend the result to open systems in the
future.

\begin{acknowledgments}
  The author thanks Innovation Program of Shanghai Municipal Education
  Commission for financial support under Grant No. 11ZZ20.
\end{acknowledgments}

\bibliographystyle{apsrev} 

\end{document}